\documentclass[aps,showpacs,prd,twocolumn,nofootinbib,nobibnotes]{revtex4-2}

\usepackage[T1]{fontenc}
\usepackage[utf8]{inputenc}
\usepackage{graphicx}
\usepackage{amssymb}
\usepackage{amsmath}
\usepackage{color}
\usepackage{float}
\usepackage[normalem]{ulem}
\usepackage{booktabs}
\usepackage{accents}
\usepackage{amsfonts}
\usepackage[colorlinks=true,
pdfstartview=FitV,linkcolor=blue,
citecolor=blue,urlcolor=blue,breaklinks=true]{hyperref}
\usepackage{array}
\usepackage{float}
\usepackage[dvipsnames]{xcolor}
\usepackage{csquotes}
\usepackage{bbold}
\usepackage{units}
\usepackage{makecell}
\usepackage{enumitem}
\usepackage{dsfont}

\usepackage{bigints}
\usepackage{amssymb}
\usepackage{systeme}
\usepackage{multirow}
\usepackage{orcidlink}

\newcolumntype{C}[1]{>{\centering\arraybackslash}m{#1}}
\renewcommand{\eqref}[1]{\mbox{Eq.~(\ref{#1})}}

\definecolor{ForestGreen}{rgb}{0.13,0.55,0.13}

\usepackage{fixmath}

\RequirePackage{color}

\begin{document}
	
	\title{Tight bounds on the Maxwell-Carroll-Field-Jackiw parameters using Fast Radio Bursts}

\author{Filipe S. Ribeiro\orcidlink{0000-0003-4142-4304}$^a$}
\email{filipe99ribeiro@hotmail.com}\email{filipe.ribeiro@discente.ufma.br}
\author{Pedro D. S. Silva\orcidlink{0000-0001-6215-8186}$^b$}
\email{pedro.dss@ufma.br}\email{pdiego.10@hotmail.com}
\author{Rodolfo Casana\orcidlink{0000-0003-1461-3038}$^{a,c}$}
\email{rodolfocasana@gmail.com}\email{rodolfo.casana@ufma.br}
\author{Manoel M. Ferreira Jr.\orcidlink{0000-0002-4691-8090}$^{a,c}$}
\email{manojr.ufma@gmail.com}\email{manoel.messias@ufma.br}
\affiliation{$^a$Programa de P\'{o}s-graduaç\~{a}o em F\'{i}sica, Universidade Federal do Maranh\~{a}o, Campus Universit\'{a}rio do Bacanga, S\~{a}o Lu\'is, Maranh\~ao 65080-805, Brazil}
\affiliation{$^{b}$Coordena\c{c}\~ao do Curso de Ci\^encias Naturais - F\'isica, Universidade Federal do Maranh\~ao, Campus de Bacabal, Bacabal, Maranh\~ao, 65700-000, Brazil}
\affiliation{$^c$Coordenação do Curso de Física Bacharelado, Universidade Federal do Maranh\~{a}o, Campus Universit\'{a}rio do Bacanga, S\~{a}o Lu\'is, Maranh\~ao 65080-805, Brazil}

\begin{abstract}

We investigate {the arrival time and the Faraday rotation} of extragalactic electromagnetic signals from fast radio bursts (FRBs) propagating through chiral cosmic media within the framework of Maxwell-Carroll-Field-Jackiw {(MCFJ)} electrodynamics. By treating the interstellar medium as a cold, ionized chiral plasma, {we derive} the time delay between two traveling signals, expressing it in terms of modified dispersion measures (DMs) {containing chiral contributions}. {The Faraday rotation angle is then} written in terms of modified rotation measures (RMs). By combining the DMs and redshift data from a set of FRBs, {we obtain} constraints on the chiral parameter magnitude at the order of $10^{-26}$--$10^{-24}$ GeV. {Using the Faraday} rotation formulae and RM measurements, {upper bounds as stringent as $10^{-43}$ GeV on the MCFJ parameters are also obtained.}
\end{abstract}

%\pacs{41.20.Jb, 11.30.Cp, 03.50.Kk, 41.90.+e, 42.25.Lc}

	\maketitle

\textbf{\textit{Introduction\label{themodel1}}} -- Fast radio bursts (FRBs) are short-duration transient events, characterized by the emission of radio {frequency pulses} with a duration of order milliseconds or less \cite{Petroff}. The first detections, performed almost 20 years ago \cite{Lorimer_FRB}, were considered singular events, but subsequently, telescopes around the Earth have observed many more. The discovery of the first repeating FRB \cite{Spitler}, named FRB 121102, confirmed that at least some of these signals could have originated from the same source. Other repeater sources have been reported \cite{Andersen}.

Radio signals traveling through the interstellar medium (ISM) are delayed {by dispersive effects produced} by the cosmic media, which is assumed to be a magnetized low-density cold plasma, whose refractive indices associated {with} right- and left-handed circularly polarized (RCP and LCP) waves,
\begin{equation}
n_{R,L}^{2}=1-{\omega_{p}^{2}}/{(\omega\left(\omega\pm\omega_{c}\right))}, \label{n_usual}
\end{equation}
lead to the respective group velocities,
\begin{equation}
v_{g}^{-1}\approx 1+\frac{\omega_{p}^{2}}{2  \omega^{2}}\pm\frac{\omega_{c}\omega_{p}^{2}}{\omega^{3}}, \label{groupvelocity}
\end{equation}
under the condition $\omega\gg\omega_{p}$. Here, $\omega_{p}=(n_{e}e^{2}/m)^{1/2}$ and $\omega_{c}=e B_{\parallel}/m$ are the plasma and cyclotron frequencies, and $e$, $m$, $n_{e}$ the electron charge, mass and number density (given in cm$^{-3}$), respectively.
The time delay is derived by taking the difference between the time of arrival, $t=\int_{0}^{d}v_{g}^{-1}ds$, for a signal traveling at the group velocity (\ref{groupvelocity}), and a wave traveling at the speed of light $c$, yielding the well-known $\omega^{-2}$ dispersion delay \cite{Lorimer-Kramer}
 \begin{equation}
 	\tau = \frac{e^{2}}{2m\omega^{2}}\mathrm{DM}.\label{timedelay}
 \end{equation}
The quantity $\mathrm{DM} = \int_{0}^{d} n_e ds$ is the dispersion measure, defined by the integral of the electron number density $n_e$ along the line of sight over the distance $d$.

Furthermore, polarization properties of the radio signals are analyzed by means of Faraday rotation, arising due to the Galactic magnetic field. The rotation angle is defined by the integration of the phase difference between circularly polarized waves,
\begin{equation}
\Delta\phi=\frac{1}{2}\int_{0}^{d}\left(k_{R}-k_{L}\right)  ds=\frac{e^{3}\lambda^{2}}{8\pi^{2}m^{2}}\int_{0}^{d}n_{e}B_{\parallel} ds, \label{Faraday}
\end{equation}
also obtained within the small density hypothesis. Usually, the Faraday angle is expressed in the form
\begin{equation}
\Delta\phi =\lambda^{2}\text{RM}, \quad \text{RM}=\frac{e^{3}}{8\pi^{2}m^{2}}\int_{0}^{d}n_{e}B_{\parallel} ds, \label{Faraday_angle}
\end{equation}
{with} RM being the rotation measure (RM), measured in $\mathrm{rad}/\mathrm{m}^{2}$. {DM and RM measurements of pulsars and galactic transient sources} have been used to investigate the ISM and the galactic magnetic field orientation \cite{Simard-Normandin}.  

Analogously, FRBs originated in other galaxies have been explored to study the intergalactic medium and the extragalactic magnetic field \cite{Vazza}, through similar analysis of their DM and RM. However, for extragalactic radio signals, redshift corrections need to be considered due to their cosmological distances. In this scenario, for a plasma at redshift $z$, the time delay (\ref{timedelay}) reads
 \begin{equation}
\tau_{z} = \frac{e^{2}}{2m\omega_{z}^{2}}\mathrm{DM}_{z},
\end{equation}
where $\mathrm{DM}_{z}= \int_{0}^{d} n_{e,z} ds$ is the rest-frame DM, which counts the column density of free electrons at the source. The distribution of free electrons in the Galaxy and other regions in the Universe is estimated by known models \cite{Yao}. In the observer frame,  the angular frequency is corrected as $\omega=\omega_{z}/(1+z)$, while the time delay becomes $\tau=\tau_{z}(1+z)$ {\cite{Wei-Deng} yielding}
\begin{equation}
\tau = \frac{e^{2}}{2m\omega^{2}}\mathrm{DM}_{\mathrm{obs}}, \qquad \mathrm{DM}_{\mathrm{obs}}=\int_{0}^{d}\frac{n_{e,z}}{1+z}ds. \label{Timedelay_redshift}
\end{equation}
Here, the subscript ``obs'' denotes the observational DM, already corrected by redshift. In general, such a quantity accounts for the DM contributions from electrons located in the Milky Way, the intergalactic medium, and the host galaxy. For instance, the influence of a hypothetical photon mass were also considered \cite{Shao} .

The redshift correction to the wavelength, rewritten as  $\lambda=\lambda_{z}(1+z)$, leads to the following rotation angle in the observer frame \cite{Akahori2}:
\begin{equation}
\Delta\phi =\lambda^{2}\text{RM}_{\mathrm{obs}}, \qquad \text{RM}_{\mathrm{obs}}=\frac{\text{RM}_{z}}{(1+z)^{2}},\label{FaradayRotation_redshift}
\end{equation}
in which $\text{RM}_{\mathrm{z}}$ represents the RM in the rest frame of the source. Eq.~(\ref{FaradayRotation_redshift}) establishes a connection between the observed RM and the host rest frame $\mathrm{RM}_{z}$, being often written as $\text{RM}_{\mathrm{source}}=\text{RM}_{\mathrm{obs}}(1+z)^{2}$ {to estimate} the magnetic fields around the source. For instance, it has been employed to investigate the FRB 121102 magneto-ionic environment \cite{Michilli} and {constrain intergalactic magnetic field \cite{Ravi}. Moreover, combined DM and RM measurements allow one to estimate the line of sight magnetic fields in the FRBs host galaxies \cite{Mannings_2}.}

Very recently, DM and RM pulsar data \cite{Pulsars} were adopted to constrain CPT-odd Lorentz-violating (LV) parameters \cite{Filipe2025}, considering a cosmic medium ruled by the Maxwell-Carroll-Field-Jackiw (MCFJ) electrodynamics \cite{CFJ}. Such a theory  is the CPT-odd part of the $U(1)$ gauge sector of the Standard Model Extension (SME) \cite{Colladay}, described (in continuous matter) by the Lagrangian density
	\begin{equation}
	\mathrm{{\mathcal{L}}}=-\frac{1}{4}G^{\mu\nu}F_{\mu\nu}    + \frac{1}{4}%
	\epsilon^{\mu\nu\alpha\beta}\left( K_{AF}\right)_{\mu}A_{\nu}F_{\alpha
		\beta}, \label{MCFJ_matter}
	\end{equation}
	in which $\left(K_{AF}\right)_{\mu}$ is the constant 4-vector that induces the Lorentz symmetry breaking, $F^{\mu\nu}$ and $G^{\mu\nu}$ are the electromagnetic field strengths in vacuum and in matter \cite{Pedroo}. This model has been investigated in several directions \cite{Urrutia,CFJG3}. In this context, the time delay (\ref{timedelay}) and the Faraday angle (\ref{Faraday_angle}) were rewritten for a chiral CFJ plasma, whose extended permittivity \cite{Filipe1},
\begin{equation}
\tilde{\varepsilon}_{ij}=\varepsilon_{ij}\left(\omega\right) + iK_{AF}^{0} \epsilon_{ikj}k^{k}/\omega+i \epsilon_{ikj}K_{AF}^{k}/\omega,\label{ChiralPlasma}
\end{equation}
involves the usual plasma dielectric tensor,
\begin{align}
\varepsilon_{ij}(\omega) &= S \delta_{ij} + i D \epsilon_{ij3} + (P -S) \delta_{i3} \epsilon_{j12},
\end{align}
with $ P = 1- {\omega_{p}^{2}}/{\omega^{2}}$, {$S=1 - {\omega_{p}^{2}}/{(\omega^{2}-\omega_{c}^{2})}$, and $D= { \omega_{c} \omega_{p}^{2}}/{\omega (\omega^{2} - \omega_{c}^{2})}$}.
In this scenario, the LV parameter magnitude was constrained at the order of $10^{-23}$ -- $10^{-22}$ GeV by using DMs data from five pulsars, while RM measurements implied upper constraints as tight as $10^{-36}$ GeV {(see Ref.~\cite{Filipe2025})}. Comparison with restrictions established in other physical scenarios was carried out,
highlighting the main constraints as for Schumann resonances (at the level of $10^{-20}$ GeV) \cite{Mewes-bounds-1}, laboratory-based analysis involving resonant cavities (at the level of $10^{-23}$ GeV) \cite{Yuri}, solar wind deviation data (as $10^{-24}$ GeV) \cite{Spallicci},  and by the anisotropies of cosmic microwave background (at {the tightest limit of $10^{-45}$} GeV) \cite{Caloni}.

%Such a systems was investigated in previous works \cite{Filipe1,Filipe2}, in which the influence of the { chiral factor} $K_{AF}^{0}$ and chiral vector %$\mathbf{K}_{AF}$ on the electromagnetic modes lead to modified effects (e.~g., birefringence and dichroism) in magnetized plasma.

%{Furthermore, the MCFJ electrodynamics provides} an effective framework to describe chiral phenomena in condensed matter, such as the chiral magnetic effect %(CME)\cite{Kharzeev1, Kharzeev1A, Kharzeev1B, Fukushima, Kharzeev1C, LiKharzeev} and anomalous Hall effect (AHE) \cite{ Haldane, Xiao, Huang, Liu}, often addressed %in Weyl semimetals. These chiral effects are connected to the MCFJ electrodynamics, with the chiral magnetic current density being written as  %$\mathbf{J}_{B}=K_{AF}^{0}\mathbf{B}$, where $K_{AF}^{0}$ plays the role of the magnetic conductivity \cite{Pedro1, PedroPRB2024A}, and the chiral vector %$\mathbf{K}_{AF}$ represents the anomalous Hall conductivity in the current $\mathbf{J}_{AH}=\mathbf{K}_{AF} \times\mathbf{E}$ \cite{Qiu}.

	%In \eqref{ChiralPlasma}, the { chiral factor} $K_{AF}^{0}$ and chiral vector $\mathbf{K}_{AF}$ lead to modified effects (e.~g., birefringence and dichroism) in %magnetized \cite{Filipe1,Filipe2} and unmagnetized plasma \cite{Filipe3}. Chiral plasma effects in astrophysics have also been explored in pulsars and black %holes --  objects surrounded by magnetospheres made of plasma -- where the CME current,  $\mathbf{J}_{B}=\mu_{5}\mathbf{B}$, is supposed to exist, with %repercussions on the propagation of helical modes \cite{Gorbar2}.

In this work, we revisit the results obtained in Ref.~\cite{Filipe2025}, including redshift corrections in order to investigate FRB radio signals. The time delay (\ref{Timedelay_redshift}) and the Faraday angle (\ref{FaradayRotation_redshift}) are rewritten in the observer frame, considering the ISM as chiral plasma described by the permittivity (\ref{ChiralPlasma}). Using FRB datasets for DMs, RMs, and redshift of some repeating and non-repeating FRBs, the additional terms are bounded in terms of the associated uncertainties on the usual ones, leading to constraints on the LV parameters. Comparisons of these new restrictions with previously studied scenarios are carried out.

\textbf{\textit{Dispersion measure in MCFJ electrodynamics}} -- {We start by} rewriting the time delay obtained in Ref.~\cite{Filipe2025}, considering redshift corrections for propagating radio waves in a chiral MCFJ plasma medium. Taking into account the timelike component in permittivity (\ref{ChiralPlasma}), the associated refractive indices \cite{Filipe1} yield the following inverse group velocity for the RCP and LCP waves for a plasma at redshift $z$:
\begin{equation}
\left(  v_{g}\right)  ^{-1}  \approx 1+\frac{\omega_{p}^{2}}%
{2\omega_{z}^{2}}+\frac{\left(K_{AF}^{0}\right)^{2}}{8\omega_{z}^{2}}.\label{groupvel_CFJ_TL}
\end{equation}
Using the redshift corrections for the angular frequency and time delay, $\omega=\omega_{z}/(1+z)$ and $\tau=\tau_{z}(1+z)$, the time delay in the observer frame becomes
\begin{equation}
\tau  =\frac{e^{2}}{2m\omega^{2}}\left[  \text{DM}_{\mathrm{obs}}+\text{DM}_{\mathrm{CFJ}}^{\left(\bullet\right)}\right], \label{Timedelay_TL}
\end{equation}
where the CFJ contribution is defined as
\begin{equation}
\text{DM}_{\mathrm{CFJ}}^{\left(\bullet\right)}=\frac{m\left(K_{AF}^{0}\right)^{2}}{4e^{2}\left(1+z \right)}d_{C},\label{DM_TL}
\end{equation}
in which $d_{C}$ denotes the {so-called \textit{comoving distance}, used for objects} at cosmological distances ({e.g.} extragalactic sources).

{For the spacelike component} in (\ref{ChiralPlasma}), the propating modes can be considered {in two} configurations, in which the chiral vector is parallel ($\mathbf{K}_{AF}\parallel \mathbf{B}_{0}$) and orthogonal ($\mathbf{K}_{AF}\perp\mathbf{B}_{0}$) to magnetic field, as investigated in Ref. \cite{Filipe1}. For these two cases, the inverse group {velocities} (at redshift $z$), associated with the RCP and LCP indices, are
\begin{align}
\left(  v_{g}\right)  ^{-1} &  \approx 1+\frac{\omega_{p}^{2}}%
{2\omega_{z}^{2}}+\frac{\lvert \mathbf{K}_{AF\parallel}\rvert^{2}}{8\omega_{z}^{2}},\label{groupvel_CFJ_SL1} \\
	\left(  v_{g}\right)  ^{-1} & \approx 1+\frac{\omega_{p}^{2}}%
	{2\omega_{z}^{2}}+\frac{\lvert \mathbf{K}_{AF\perp}\rvert^{2}}{2\omega_{z}^{2}},\label{groupvel_CFJ_SL2}
\end{align}
respectively, {yielding the time delays} (in the observer frame) 
\begin{align}
\tau  =\frac{e^{2}}{2m\omega^{2}}\left[  \text{DM}_{\mathrm{obs}}+\text{DM}_{\mathrm{CFJ}}^{\left(\bullet\bullet\right)}\right], \label{Timedelay_SL} \\
\tau  =\frac{e^{2}}{2m\omega^{2}}\left[  \text{DM}_{\mathrm{obs}}+4\text{DM}_{\mathrm{CFJ}}^{\left(\bullet\bullet\right)}\right], \label{Timedelay_SL2}
\end{align}
where
\begin{equation}
\text{DM}_{\mathrm{CFJ}}^{\left(\bullet\bullet\right)}=\frac{m\lvert \mathbf{K}_{AF\parallel}\rvert^{2}}{4e^{2}\left(1+z \right)}d_{C}. \label{DM_SL}
\end{equation}
Note that the time delay (\ref{Timedelay_SL}) and the CFJ contribution (\ref{DM_SL}) present the same form as the one of the timelike case in (\ref{Timedelay_TL}) and (\ref{DM_TL}). Comparing with the previous results {of Ref.~\cite{Filipe2025}}, the cosmological correction {to the} additional contributions $\mathrm{DM}_{\mathrm{CFJ}}$ in expressions (\ref{Timedelay_TL}), (\ref{Timedelay_SL}), and (\ref{Timedelay_SL2}), replaces the usual distance $d$ by the {comoving distance} as $d\rightarrow d_{C}/(1+z)$, {yielding a} redshift-dependent chiral contribution.

\textbf{\textit{Faraday rotation MCFJ electrodynamics}} -- Exploring now the Faraday effect, we take the RCP and LCP wave vectors {associated with the indices} obtained for a timelike and spacelike CFJ plasma \cite{Filipe1} {(for the $\mathbf{K}_{AF} \parallel \mathbf{B}$ case)}, under the low-density condition and at redshift $z$, as
 \begin{align}
 k_{R,L} &\approx \Omega(z) \mp \frac{K_{AF}^{0}}{2}+\frac{\left(K_{AF}^{0}\right)^{2}}{8\omega_{z}}, \\
  k_{R,L} & \approx \Omega(z) \mp \frac{\lvert \mathbf{K}_{AF\parallel}\rvert}{2}-\frac{\lvert \mathbf{K}_{AF\parallel}\rvert^{2}}{8\omega_{z}}\mp \frac{\omega_{p}^{2}\lvert \mathbf{K}_{AF\parallel}\rvert}{4\omega_{z}^{2}},
 \end{align}
where {$\Omega(z)=\omega_{z}-\omega_{p}^{2}/(2\omega_{z}) \pm 
\omega_{c}\omega_{p}^{2}/(2\omega_{z}^{2})$}. In the observer frame, using $\lambda=\lambda_{z}(1+z)$, such wave vectors yield modified Faraday rotation expressions 
 \begin{align}
 \Delta\phi  =\lambda^{2}\left(\mathrm{RM_{obs}}-\mathrm{RM_{CFJ}}^{\left(\bullet\right)}\right), \label{FRB_RM_1} \\
  \Delta\phi  =\lambda^{2}\left(\mathrm{RM_{obs}}-\mathrm{RM_{CFJ}}^{\left(\bullet\bullet\right)}\right),\label{FRB_RM_2}
 \end{align}
where $\mathrm{RM_{obs}}$ is observational RM given in (\ref{FaradayRotation_redshift}), and the second terms are $\lambda$-dependent CFJ contributions,
 \begin{align}
 \mathrm{RM_{CFJ}}^{\left(\bullet\right)} & =\frac{K_{AF}^{0}}{2\lambda^{2}}d_{C}\label{RM_CFJ_1}, \\
 \mathrm{RM_{CFJ}}^{\left(\bullet\bullet\right)} & =\frac{\lvert \mathbf{K}_{AF\parallel}\rvert}{2\lambda^{2}}d_{C}+\frac{e^{2}\lvert \mathbf{K}_{AF\parallel}\rvert}{4\kappa\left(1+z\right) }\mathrm{DM_{obs}},\label{RM_CFJ_2}
\end{align}
with $\kappa=4\pi^{2}m$. By comparing with the results in Ref.~\cite{Filipe2025}, {we note} that $\mathrm{RM_{CFJ}}^{\left(\bullet\bullet\right)}$ {exhibits an explicit redshift correction}, while $\mathrm{RM_{CFJ}}^{\left(\bullet\right)}$ {is modified only by the comoving distance} $d_{C}$.

\textbf{\textit{Constraints using Fast Radio Bursts}} -- Here, we establish constraints on the LV parameters using observational dispersion and rotation measure data from FRBs. In our framework, the additional contributions $\mathrm{DM_{CFJ}}$ and $\mathrm{RM_{CFJ}}$, given in Eqs. (\ref{Timedelay_TL}), (\ref{Timedelay_SL}), (\ref{Timedelay_SL2}), (\ref{FRB_RM_1}) and (\ref{FRB_RM_2}), are treated as small corrections bounded by the observational uncertainties on $\mathrm{DM_{obs}}$ and $\mathrm{RM_{obs}}$, denoted by $\epsilon_{\text{DM}}$ and $\epsilon_{\text{RM}}$, respectively. In this sense, the constraints will be determined from
\begin{equation}
\mathrm{DM_{CFJ}}\lesssim\epsilon_{\text{DM}}, \qquad \mathrm{RM_{CFJ}}\lesssim\epsilon_{\text{RM}}.
\end{equation}
We will consider data {from three} repeating FRBs, namely FRB 121102, FRB 20180916B, and FRB 20190303A, and four non-repeating FRBs, FRB 20190102, FRB 20190608, FRB 191108, and FRB 20220610A. In this procedure, the {comoving distance} $d_{C}$ is determined by cosmological parameters from the Planck data \cite{Planck18}, by using the Python package \texttt{astropy.cosmology} \cite{Astropy}.

\textit{DM constraints using FRBs} -- Starting by expression (\ref{Timedelay_TL}), the timelike chiral contribution in time delay, given in Eq. (\ref{DM_TL}), is bounded as
\begin{equation}
K_{AF}^{0}\lesssim 2e\sqrt{\frac{\left(1+z \right)\epsilon_{\mathrm{DM}}}{md_{C}}}, \label{DM_Constraint_TL}
\end{equation}
from which we can derive restrictions on the chiral factor $K_{AF}^{0}$ using the conventional values for the electron charge and mass. Here, we take data of two non-repeating FRBs, FRB 20190102 and FRB 20190608, provided by the Australian Square Kilometre Array Pathfinder (ASKAP) \cite{Day}, whose associated observational DM is $\mathrm{DM}=(364.545 \pm 0.004)$ pc $\mathrm{cm}^{-3}$ and $\mathrm{DM}=(340.05 \pm 0.06)$ pc $\mathrm{cm}^{-3}$, respectively.
In addition, the respective comoving distances can be determined by the redshifts available in Ref.~\cite{Mannings}, providing $z=0.2912$ ($d_{C}\approx1199.3$ Mpc) and $z=0.1177$ ($d_{C}\approx506.9$ Mpc), respectively. Now, inserting {the DM uncertainties for FRB 20190102 and FRB 20190608, $\epsilon_{\text{DM}}=0.004$ pc $\mathrm{cm}^{-3}$ and $\epsilon_{\text{DM}}=0.06$ pc $\mathrm{cm}^{-3}$, into condition (\ref{DM_Constraint_TL}) yields}
\begin{align}
K_{AF}^{0}\lesssim 1.5 \times 10^{-25}~\mathrm{GeV},\\
K_{AF}^{0}\lesssim 8.5 \times 10^{-25}~\mathrm{GeV},
\end{align}
respectively, for these two FRBs. The above constraints are also valid for the CFJ vector $\mathbf{K}_{AF}$ {(for the case $\mathbf{K}_{AF}\parallel \mathbf{B}_{0}$)}, due to the similarity between the chiral DMs in (\ref{DM_TL}) and (\ref{DM_SL}). For the case $\mathbf{K}_{AF}\perp\mathbf{B}_{0}$, the modified time delay (\ref{Timedelay_SL2}) implies $4\mathrm{DM}_{CFJ}\lesssim\epsilon_{\text{DM}}$, resulting in
\begin{equation}
\lvert \mathbf{K}_{AF}\rvert\lesssim e\sqrt{\frac{\left(1+z \right)\epsilon_{\mathrm{DM}}}{md_{C}}} .\label{DM_Constraint_SL}
\end{equation}
In this case, the data for the FRB 20190102 and FRB 20190608 imply the restrictions
\begin{align}
		\lvert \mathbf{K}_{AF}\rvert\lesssim 7.6 \times 10^{-26}~\mathrm{GeV},  \\ \label{DM_Constraint_SL_2}	
		\lvert \mathbf{K}_{AF}\rvert\lesssim 4.2 \times 10^{-25}~\mathrm{GeV}. 
\end{align}
We further consider other four FRBs, 20110220, 20110627, 20110703, and 20120127, whose data {(available in Ref.~\cite{Thornton})} are used to derive constraints on both timelike and spacelike parameters in the order of $10^{-25}$ -- $10^{-24}$ GeV. Table \ref{tab:table_constraints_nonrepeating_FRB_DM} summarizes these results, representing a two orders of magnitude improvement {compared with} the DM LV bounds previously obtained with pulsars \cite{Filipe2025}, established in order of $10^{-23}$ GeV. 
	\begin{table}[htbb]
		\centering
		\caption{Constraints on LV parameters using DM for FRBs.}
		\label{tab:table_constraints_nonrepeating_FRB_DM}
		
		\renewcommand{\arraystretch}{1.2}
		\setlength{\tabcolsep}{5pt}
		
		\begin{tabular}{*{4}{c}}
			\toprule[0.8pt]\midrule
			\textbf{FRBs} & \makecell{$\mathrm{DM}_{\mathrm{obs}}$\\ ($\mathrm{pc \,cm}^{-3}$)	 } &  \makecell{$K_{AF}^{0}$ and $\mathbf{K}_{AF}^{\parallel}$\\ (GeV)} & \makecell{$\mathbf{K}_{AF}^{\perp}$\\ (GeV)}  \\
			\midrule
		20190102 &  $364.545 \pm 0.004$   &      $1.5\times 10^{-25}$ &  $7.6\times 10^{-26}$ \\
	    20190608  & $340.05 \pm 0.06$ &    $8.5 \times 10^{-25}$ & $4.2 \times 10^{-25}$ \\
		 20110220  & $944.38 \pm 0.05$ &    $4.2 \times 10^{-25}$ & $2.1 \times 10^{-25}$ \\ 	
		20110627  & $720.0 \pm 0.3$ &    $1.1 \times 10^{-24}$ & $5.5 \times 10^{-25}$ \\ 	
		20110703  & $1103.6 \pm 0.7$ &    $1.5 \times 10^{-24}$ & $6.6 \times 10^{-25}$ \\
		20120127  & $553.3 \pm 0.3$ &    $1.3 \times 10^{-24}$ & $6.9 \times 10^{-25}$ \\
			\midrule\bottomrule[0.8pt]
		\end{tabular}
	\end{table}

\textit{RM constraints using repeating FRBs} -- Following the same procedure used for DM restrictions, we can limit the LV parameter magnitude employing the RM uncertainty as an upper limit of the RM quantities given in Eqs. (\ref{RM_CFJ_1}) and (\ref{RM_CFJ_2}), attaining
\begin{align}
K_{AF}^{0} & \lesssim\frac{2\lambda^{2}}{d_{C}}\epsilon_{\text{RM}}, \label{FRB_contraint_1} \\
\lvert\mathbf{K}_{AF}\rvert & \lesssim\frac{8\lambda^{2}\kappa\left(1+z\right)}{4\kappa\left(1+z\right) d_{C}+2\lambda^{2}e^{2}\mathrm{DM}}\epsilon_{\text{RM}} \label{FRB_contraint_2}.
\end{align}
We then follow our analysis with the repeating FRB 121102, {using DM and RM data from} Ref.~\cite{Hilmarsson}. {We select} some bursts observed with Arecibo and Effelsberg telescopes, with centered frequencies around 4.5 GHz and 6 GHz, ($\lambda\approx0.0666205~\mathrm{m}$) and ($\lambda\approx0.0499654~\mathrm{m}$), respectively.  Moreover, the associated redshift, $z=0.193$, yields a {comoving distance} $d_{C}\approx 814.4$~Mpc.
Taking into account the above distance and the data associated with the bursts 10, 13, 18, and 19 (see Table I in Ref.~\cite{Hilmarsson}), the conditions   (\ref{FRB_contraint_1}) and (\ref{FRB_contraint_2}) stablish constraints of the order of $10^{-43}$ and $10^{-42}$ GeV for both timelike and spacelike parameters, as presented in the four top rows in {Tab.~\ref{tab:table_constraints_repeating_FRB}}.

{We also consider} the repeating FRB 20180916B, localized to an edge of aspiral galaxy at a redshift of 0.0337 ($d_{C}\approx 148.1$~Mpc), with data provided in Ref.~\cite{Bethapudi}, from which {we select} four bursts observed around 650 MHz ($\lambda\approx0.461219~\mathrm{m}$). From the {third repeating} source, FRB 20190303A, localized at redshift $z=0.064$ (with $d_{C}\approx279.2$ Mpc), {we use data} from three bursts observed around 700 MHz ($\lambda\approx0.428275 ~\mathrm{m}$) \cite{Mckinven}. These two set of data, by restrictions (\ref{FRB_contraint_1}) and (\ref{FRB_contraint_2}), yield constraints in the order of $10^{-41}$ GeV and $10^{-42}$ on the LV parameters, respectively, as we can see in the last two slots of rows in {Tab.~\ref{tab:table_constraints_repeating_FRB}}. It represents an {improvement} by a factor of $10^{7}$ in comparison to the RM bounds obtained from pulsar signals \cite{Filipe2025}.

\textit{RM constraints using non-repeating FRBs} -- {We collect} data from three sources:

$i)$ The non-repeating FRB 20190102 and FRB 20190608 were observed by the Australian Square Kilometre Array Pathfinder (ASKAP) telescope at frequency reference of 1.3 GHz \cite{Mannings}, from which data on DM and RM are presented in {Tab.~\ref{tab:table_constraints_nonrepeating_FRB}}. {Their associated redshifts} \cite{Day}, given by $z=0.2912$ and $z=0.1177$, imply comoving distances $d_{C}\approx1199.3$ Mpc and $d_{C}\approx506.9$ Mpc, respectively. 
	
$ii)$ The FRB 191108 was detected with Apertif Radio Transient System (ARTS) on the Westerbork Synthesis Radio Telescope (WSRT) \cite{Connor}, {with DM and RM presented in Tab.~\ref{tab:table_constraints_nonrepeating_FRB}}, with centered frequency of 1370 MHz. We use the provided maximum redshift $z=0.52$ ($d_{C}\approx2013.2$ Mpc).

$iii)$ The FRB 20220610A, localized in a host galaxy at a redshif $1.016$, with the longest {comoving distance} considered in this work - $d_{C}\approx3435.2$ Mpc. The associated DM and RM were obtained at a centered frequency of 1271.5 MHz by ASKAP \cite{Ryder}.

\begin{widetext}	
	%usando um espaço (linha em branco) entre o comando \begin{widetext} and the \begin{table}, o titulo da tabela ira ocupar toda a largura da página.

	\begin{table}[htbb]
		\centering
		\caption{Constraints on CPT-odd Lorentz violation parameters from repeating FRB observations.}
		\label{tab:table_constraints_repeating_FRB}
		
		    \renewcommand{\arraystretch}{1.2}
		\setlength{\tabcolsep}{13pt}
		
		\begin{tabular}{*{5}{c}}
			\toprule[0.8pt]\midrule
			\textbf{FRBs} &  $\mathrm{DM}_{\mathrm{obs}}$ ($\mathrm{pc \,cm}^{-3}$)	&  $\mathrm{RM}_{\mathrm{obs}}$ (rad $\mathrm{m^{-2}}$) & $K_{AF}^{0}$ and $\mathbf{K}_{AF}$ (GeV) & Selected burts   \\
			\midrule
			FRB 121102  &  $ 561.6$ & $72248(21)$ &  $8.2\times 10^{-43}$ &  \multirow{4}{*}{\shortstack{Bursts 10, 13,\\
				18, and 19 \\from Ref.~\cite{Hilmarsson}. } }  \\
		FRB 121102  &  $561.7$ & $71525(3)$ &  $2.1\times 10^{-43}$ &   \\
			FRB 121102  & $563.2$   & $69408(92)$    &   $6.4\times 10^{-42}$ &   \\
	    FRB 121102  & $563.1$   & $66949(11)$    &   $7.6\times 10^{-43}$ &   \\ \hline
	     FRB 20180916B  &   $348.82$  & $-121.12 \pm 1.74$ &    $3.2\times 10^{-41}$ &  \multirow{4}{*}{\shortstack{Bursts 1, 4,\\
	     		5, and 8 \\from Ref.~\cite{Bethapudi}. } }  \\
		 FRB 20180916B  &   $ 348.82$  & $-115.22 \pm 0.68$ &    $1.9\times 10^{-41}$ &   \\
		
			 FRB 20180916B  &  $ 348.82$  & $-111.34 \pm  3.5$ &    $6.4\times 10^{-41}$ &  \\
		
		 FRB 20180916B  &  $ 348.82$  & $-116.48 \pm  3.7$ &    $6.8\times 10^{-41}$ &   \\ \hline
		
		  FRB 20190303A &  $ 221.33$  & $-362.65(73)$ &    $6.1\times 10^{-42}$ &  \multirow{3}{*}{\shortstack{Bursts 14, 15,\\
		  		and 16 \\from Ref.~\cite{Mckinven} }. } \\
		
		  FRB 20190303A  &  $221.78$  & $-408.76(33)$ &    $2.7\times 10^{-42}$ &   \\
		
		  FRB 20190303A  &  $ 221.72$  & $-421.33(37)$ &    $3.1\times 10^{-42}$ &   \\
			\midrule\bottomrule[0.8pt]
		\end{tabular}

	\end{table}	
\end{widetext}

For these four non-repeating sources, Eqs.~(\ref{FRB_contraint_1}) and (\ref{FRB_contraint_2}) yield constraints as tight as $10^{-43}$ GeV. See the fourth column in Tab.~\ref{tab:table_constraints_nonrepeating_FRB}. It also represents an improvement of 7 orders of magnitude over the best bounds obtained with pulsar RM \cite{Filipe2025}.
\begin{widetext}	

	%usando um espaço (linha em branco) entre o comando \begin{widetext} and the \begin{table}, o titulo da tabela ira ocupar toda a largura da página.
	
	\begin{table}[htbb]
		\centering
		\caption{Constraints on CPT-odd Lorentz violation parameters from non-repeating FRB observations.}
		\label{tab:table_constraints_nonrepeating_FRB}
		
		\renewcommand{\arraystretch}{1.2}
		\setlength{\tabcolsep}{13pt}
		
		\begin{tabular}{*{6}{c}}
			\toprule[0.8pt]\midrule
			\textbf{FRBs} & $z$& $\mathrm{DM}_{\mathrm{obs}}$ ($\mathrm{pc \,cm}^{-3}$)	&  $\mathrm{RM}_{\mathrm{obs}}$ (rad $\mathrm{m^{-2}}$) & $K_{AF}^{0}$ and $\mathbf{K}_{AF}$ (GeV) & Ref.   \\
			\midrule
			FRB 20190102  & $0.2912$&  $364.545$ & $42.8\pm 0.2$ &  $1.1\times 10^{-43}$ &  \cite{Day,Mannings} \\
			FRB 20190608 & $0.1177$&  $340.05$   &  $42.3\pm 0.1$     &   $1.3\times 10^{-43}$ & \cite{Day,Mannings} \\
		FRB 20191108  &  $0.52$&  $ 588.1$  & $474 \pm  3$ &    $6 \times 10^{-43}$ & \cite{Connor} \\
		FRB 20220610A  &  $1.016$&  $  1458.15$  & $215 \pm  2$ &    $5\times 10^{-43}$ & \cite{Ryder} \\
			\midrule\bottomrule[0.8pt]
		\end{tabular}
		
	\end{table}	
\end{widetext}

\textbf{\textit{{Final remarks\label{conclusion}}}} -- In this letter, we have employed FRB observations to {constrain} \textit{CPT}-odd Maxwell-Carroll-Field-Jackiw parameters. Assuming the interstellar medium as a chiral plasma, {we have derived} modified expressions for both the time delay and Faraday rotation angle in the observer frame, inserting cosmological corrections compatible with extragalactic radio sources. We have considered data for some repeating and non-repeating FRBs, {using} their respective DM, RM, and redshift. The modified time delay is expressed in terms of additional chiral contributions $\mathrm{DM_{CFJ}}$, given in (\ref{DM_TL}) and (\ref{DM_SL}). These chiral DMs are considered to restrain the parameters $K_{AF}^{0}$ and $\lvert \mathbf{K}_{AF}\rvert$ (for parallel configuration) according to Eq.~(\ref{DM_Constraint_TL}), while $\lvert \mathbf{K}_{AF}\rvert$  (for orthogonal configuration) is bounded by Eq.~(\ref{DM_Constraint_SL}). By limiting the chiral contribution to the observed DM uncertainties of the non-repeating FRBs 20190102 and 20190608, {we have obtained} constraints in the order of $10^{-26}-10^{-24}$ GeV, as shown in Tab.~\ref{tab:table_constraints_nonrepeating_FRB_DM}.  

{Moreover}, a significant improvement {arises} when considering data from Faraday rotation measures, for which the chiral parameters are bounded by Eqs. (\ref{FRB_contraint_1}) and (\ref{FRB_contraint_2}). In this case, we have taken data of three repeating sources, FRBS 122202, 20180916B, and 20190303A, for which we obtained constraints as severe as $10^{-41}-10^{-43}$ GeV, as presented in Tab.~\ref{tab:table_constraints_repeating_FRB}. Four non-repeating sources, FRBs 20190102, 20190608, 20191108, and 20220610A, were also considered, yielding {constraints of order} $10^{-43}$ GeV. Such results improve the {analogous bounds} obtained from pulsars RM \cite{Filipe2025} by a factor of $10^{7}$, representing a relevant enhancement.

 Table~\ref{tab:table_constraints3} compares the present results with others {established} in different contexts. We can see that the constraints using RM of FRBs are more restrictive than the bounds obtained from solar wind, pulsars DM, resonant cavities, and Schumann resonances. Although the {most restrictive} results remain from CMB polarization, {the bounds} from FRB RM are competitive to the best astrophysical birefringence bounds in the literature, strengthening the actual interest for investigating chiral plasmas in ISM.
 	
\begin{table}[H]
	\centering
	\caption{{Bounds comparison among distinct setups.}}
	\label{tab:table_constraints3}
	\begin{tabular}{*{4}{c}}
		\toprule[0.8pt]\midrule
		\textbf{---} &  $K_{AF}^{0}$ (GeV)	&  $\mathbf{K}_{AF}$ (GeV) & Ref.   \\
		\midrule
		{CMB polarization} & $10^{-45}$ & {$10^{-45}$} &   \cite{Caloni} \\
		{\textbf{Fast Radio Bursts RM}} & $10^{-43}$  & $10^{-43}$ &   \textbf{Table} \ref{tab:table_constraints_nonrepeating_FRB} \\
		{Astrophys. Birefringence} & $10^{-42}$  & $10^{-42}$ &   \cite{CFJ} \\
		Pulsar RM & $10^{-36}$ & $10^{-36}$ &  \cite{Filipe2025} \\
			{\textbf{Fast Radio Bursts DM}} & $10^{-25}$  & $10^{-25}$ &   \textbf{Table} \ref{tab:table_constraints_nonrepeating_FRB_DM} \\
		Solar wind &  -  & $10^{-24}$ &   \cite{Spallicci} \\
	Pulsar DM  &  $10^{-22}$ &  $10^{-23}$ &  \cite{Filipe2025}\\
		Resonant cavities &   -- &  $10^{-23}$ &    \cite{Yuri} \\	
		Schumann resonances & $10^{-21}$   & $10^{-21}$   &   \cite{Mewes-bounds-1} \\
		\midrule\bottomrule[0.8pt]	
	\end{tabular}
\end{table}

\textit{\textbf{Acknowledgments}}. The authors thank FAPEMA, CNPq, and CAPES (Brazilian research agencies) for their invaluable financial support. M.M.F. is supported by FAPEMA APP-12151/22, CNPq/Produtividade 317048/2023-6, and CNPq/Universal/420896/2025-2. P.D.S.S. is grateful to FAPEMA APP-12151/22 and Produtividade/FAPEMA/CNPq/12649/25. Furthermore, we are indebted to CAPES/Finance Code 001 and FAPEMA/POS-GRAD-04755/24.

\end{document}